\documentclass[twocolumn,showpacs,amsmath,amssymb,floatfix,pre]{revtex4}
\usepackage{graphics,graphicx,epsfig}

\begin{document}
\title{Roughness exponents and grain shapes}
\author{T. J. Oliveira${}^{1,(a)}$ and F. D. A. Aar\~ao Reis${}^{2,(b)}$
\footnote{a) Email address: tiago@ufv.br\\
b) Email address: reis@if.uff.br}}
\affiliation{ ${}^{1}$ Departamento de F\'isica, Universidade Federal de Vi\c cosa,
36570-000, Vi\c cosa, MG, Brazil \\
${}^{2}$ Instituto de F\'\i sica, Universidade Federal
Fluminense, Avenida Litor\^anea s/n, 24210-340 Niter\'oi RJ, Brazil}

\date{\today}

\begin{abstract}

In surfaces with grainy features, the local roughness $w$
shows a crossover at a characteristic length $r_c$, with
roughness exponent changing from $\alpha_1\approx 1$ to a smaller $\alpha_2$..
The grain shape, the choice of $w$ or height-height correlation function
(HHCF) $C$,
and the procedure to calculate root mean-square averages are shown to have
remarkable effects on $\alpha_1$. With grains of pyramidal shape, $\alpha_1$
can be as low as $0.71$, which is much lower than the previous prediction
$0.85$ for rounded grains.
The same crossover is observed in the HHCF, but with initial exponent
$\chi_1\approx 0.5$ for flat grains, while 
for some conical grains it may increase to $\chi_1\approx 0.7$.
The universality class of the growth process determines the
exponents $\alpha_2=\chi_2$ after the crossover, but
has no effect on the initial exponents $\alpha_1$ and $\chi_1$,
supporting the geometric interpretation of their values.
For all grain shapes and different definitions of surface roughness or HHCF,
we still observe that the crossover length $r_c$ is an accurate estimate of the grain size.
The exponents obtained in several recent experimental works on different materials are
explained by those models, with some surface images qualitatively similar to our
model films.

\end{abstract}

\maketitle

\section{Introduction}
\label{intro}

Scaling properties of the local surface roughness $w$ and of the height-height correlation
function (HHCF) $C$ are very useful to understand the growth dynamics of thin films and
other deposits \cite{barabasi,krug,krim}. The usual approach is to measure exponents
from plots of $w$ or $C$ as a function of the box size $r$ (roughness exponent) or time $t$
(growth exponent) and to relate their values
to some universality class of growth \cite{barabasi}.
However, a very small number of systems exibit simple scaling features to match those
theories. For instance, the presence of grains in the film surface leads to a crossover
between two regimes where $w$ increases with $r$ with different roughness exponents,
$\alpha_1$ and $\alpha_2$, as illustrated in Fig. 1 
\cite{lita,kleinke1999,ebothe2003,vazquez,mendez,otsuka,vasco}.
For the HHCF, the same crossover occurs with exponents $\chi_1$ and $\chi_2$..
Similar crossover is observed in other systems, such as fresh snow on the ground and
pyroclastic deposits on volcanic surfaces \cite{manes,mazzarini}.

In Ref.\protect\cite{grains}, the crossover with $\alpha_1\approx 1$
was shown to be a geometric effect of the grainy surface structure and of the gliding box method
(analogous result is obtained with the box counting method).
It was also shown that the crossover took place when $r$ was close to the average grain size.
If the grain surface is flat, $\alpha_1$ is very close to $1$, while for rounded grains
it decreases to values close to $0.85$ \cite{grains}. These results match
those of a large number of experimental works
\cite{lita,kleinke1999,ebothe2003,vazquez,mendez,otsuka,vasco}.
However, other experimental works show film surfaces with grainy structure, the same
crossover in roughness scaling or HHCF, but with much smaller exponents $\alpha_1$
\cite{vazquezSS,tersio,mendomeJCG,mendomeMCP,hiane,nara2005,nara2007,naraNano2007,marta2002}.
The usual interpretation for those exponents is that small
scale surface features are determined by a different growth dynamics.
Indeed, even the crossover with $\alpha_1\approx 1$ was already interpreted as an
anomalous scaling, with $\alpha_1$ being called local roughness exponent (denoted
$\alpha_{loc}$) and $\alpha_2$ called global roughness exponent. For these reasons,
in many systems it is still unclear whether a crossover similar to that in Fig. 1
should be interpreted as a purely geometric effect or as a consequence of a
competitive growth dynamics. 

\begin{figure}[!b]
\includegraphics[width=7cm]{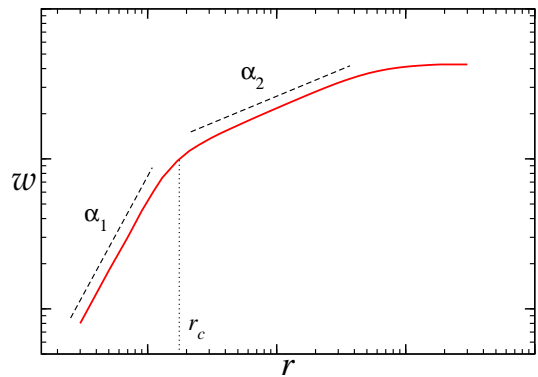}
\caption{Typical behavior of the local roughness as a function of box size in grainy surfaces.}
\label{fig1}
\end{figure}

Here we study several growth models with grainy surface features to show the
possible effects of the grain shape, of the method of calculation of
averages of squared quantities, of the working quantity ($w$ or $C$) and
of the universality class of the growth process.
For all growth models, grain shapes, and methods of analysis, we observe
crossovers at box sizes very close to the average grain size.
We also show that a very broad range of $\alpha_1$ can be found, depending on the
grain shape and the working quantity, but independently of the universality class of growth,
which determines only the value of $\alpha_2$. Similar conclusions are obtained for the
exponents $\chi_1$ and $\chi_2$. The comparison with experimental works with
several materials and deposition methods gives additional support to the geometric
interpretation of the crossover in those systems.

The rest of this work is organized as follows. In Sec. \ref{model} we define average quantities
and present the growth models. In Sec. \ref{theoretical}, we recall the results of some
exactly solvable models with grains at the surface, which explain the crossover with
$\alpha_1\approx 1$ and $\chi_1\approx 0.5$ (with the usual definition of the HHCF).
In Sec. \ref{shape} we analyze the effects
of the grain shape, particularly some very sharp grains, considering models in
different universality classes. In Sec. \ref{experimental}, we show the applications
of our approach to real films. In Sec. \ref{conclusion}, we summarize our
results and present our conclusions.

\section{Definition of average quantities and models}
\label{model}

First we define the average quantities analyzed in this work.

The surface roughness in square boxes of size $r$ at time $t$ is usually defined as
\begin{equation}
w{\left( r,t\right)}\equiv {\left<
{\overline{{\left( h-\overline{h}\right)}^2}}^{1/2}\right>} .
\label{defw}
\end{equation}
The overbars in Eq. (\ref{defw}) denote averages of the height $h$ inside a given
box position (spatial average)
and the angular brackets represent the configurational average as the box
scans the whole surface of a deposit. This is called gliding box method, in which the scanning
box moves one pixel each time it performs a new spatial average.
In box counting methods, the surface is divided in nonintersecting boxes for the
configurational average.

Alternatively, some authors define the roughness as
\begin{equation}
w' \equiv {\left< \overline{{ \left( h -\overline{h}\right) }^2} \right>}^{1/2} ,
\label{defw1}
\end{equation}
i. e. they calculate the configurational average of the square height fluctuation and
take the square root of that average.

When several images of a deposit are available, or several configurations are grown with the
same model, these different samples also contribute to the above configurational averages.

For window sizes below the grain size, the roughness scales as
\begin{equation}
w{\left( r,t\right)} \sim r^{\alpha_1} ,
\label{defalpha1}
\end{equation}
which defines the initial roughness exponent $\alpha_1$ (Fig. 1).

The height-height correlation function (HHCF) at distance $r$ and time $t$ is usually defined as
\begin{equation}
C\left( r,t\right) \equiv {\langle {\left[ h\left( r_0+r,t\right) -h\left( r_0,t\right)
\right] }^2\rangle}^{1/2} ,
\label{defcorr}
\end{equation}
with configurational averages taken over all different initial positions $r_0$.
Alternatively, it can be defined as
\begin{equation}
C'\left( r,t\right) \equiv {\langle {| h\left( r_0+r,t\right) -h\left( r_0,t\right) |}
\rangle} ,
\label{defcorr1}
\end{equation}
which corresponds to an interchange of the
configurational average and the calculation of the square root
in Eq. (\ref{defcorr}). In this sense, the calculation of $C(r,t)$
parallels that of $w'$, while the calculation of $C'(r,t)$ parallels that of $w$.

For window sizes below the grain size, the HHCF scales as
\begin{equation}
C{\left( r,t\right)} \sim r^{\chi_1} ,
\label{defchi1}
\end{equation}
which defines the initial roughness exponent $\chi_1$ for that function.

For window sizes much larger than the grain size (i. e. $r\gg r_c$ - see Fig. 1),
a surface obeying normal scaling has
$w\sim r^{\alpha_2}$ and $C\sim r^{\chi_2}$, with $\alpha_2=\chi_2$.
The quantities $w'$ and $C'$ obey the same scaling. Those exponents are
representative of the large lengthscale kinetics governing the growth process. Typical
examples of growth kinetics are those of Edwards-Wilkinson (EW)
\cite{ew}, of Kardar-Parisi-Zhang (KPZ) \cite{kpz},
and the diffusion-dominated ones, linear (Mullins-Herring - MH) \cite{mh} or
nonlinear (Villain-Lai-Das Sarma - VLDS) \cite{villain,laidassarma}.

Now we present the models for growth of thin films with grains at the surface.

Intrinsic corrections to scaling for large $r$ and large $t$ should be avoided in those
models, so that any crossover is solely due to the grainy structure.
This request excludes the grain deposition models
introduced in Ref. \protect\cite{grains} and related ballistic-like models
\cite{bbd,intrinsic} because they have remarkable scaling corrections.
On the other hand, some models with smooth surfaces and particle enlargement presented in
Refs. \protect\cite{vazquez,grains} satisfy that condition. They are described below.

The first model has KPZ kinetics. The first step is to grow a deposit with cubic particles
of unit size
following the rules of the restricted solid-on-solid (RSOS) model: the aggregation
of the incident particle is accepted only if the height differences of nearest neighbors
are always $0$ or $1$ (otherwise the aggregation attempt is rejected) \cite{kk}.
We recall that $\alpha_2=\chi_2\approx 0.39$ for the KPZ class in
two-dimensional substrates \cite{eqkpz}.

The second model has VLDS kinetics. The initial deposit is grown with the rules of the
conserved RSOS model, where the incident particle executes a random walk between
neighboring columns until finding a column where it can aggregate respecting the conditions
on height differences \cite{crsoskim,crsos}.
We recall that $\alpha_2=\chi_2\approx 0.67$ for the VLDS
class in two-dimensional substrates \cite{janssen,crsos}.

\begin{figure}[!h]
\includegraphics[width=9cm]{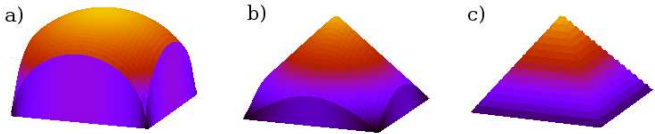}
\caption{Shapes of surface grains after enlargement of the original deposits:
a) semi-elliptical, b) conical, c) pyramidal.
Semi-ellipsoids and cones are cut at the sides so that their basis are squares
of side $l$ that fit the shape of the cubic grain at their bottom.
This is the reason for radius $l\sqrt{2}/2$ of their basis.}
\label{fig2}
\end{figure}

After growing the initial deposit, with KPZ or VLDS model,
the size of each particle is enlarged by a factor $l$, i. e. each particle is
transformed in a cubic grain of side $l$. Most of our simulations are performed with
$l=32$. 
The final step is replacing the top cub grains (surface grains) by rounded or sharp
structures. Three shapes are used:
semi-ellipsoids of horizontal radius $l\sqrt{2}/2$ and vertical radius $h$,
cones with that radius and height $h$, and pyramids of square basis of side $l$
and height $h$. They are illustrated in Fig. \ref{fig2}. Several values of $h$ are considered
for each shape, typically between $l$ and $3l$.

In the scaling of $w$ or $C$, the role of the height $h$
is measured relatively to the height of the surface steps, which is $l$. The
horizontal scaling factor is also $l$ for the cubic grains, but this is not important
for the scaling exponents. For instance, if the grains were constructed with the shape
of paralelepids of height $l$ and horizontal sides $l_{\|}$,
the scaling exponents would not change.
Thus, the aspect ratio of the grains considered here is not a limitation of the model.

The simulations of the KPZ and VLDS models
were performed in square substrates (three-dimensional
deposits) of lateral size $L=128$ at times of order ${10}^4$. For the RSOS model, it
corresponds to approximately $5\times {10}^3$ layers of unit size particles; for the
CRSOS model, corresponds to ${10}^4$ layers. After replacement of the original particles
by grains of size $l=32$, the deposits have lateral size $4096$.
Simulations in smaller sizes ($L=64$ and $L=32$ for the original models)
and different grain size ($l=16$) give similar results for all exponents,
indicating that finite-size and finite-time effects are negligible.

\section{Theoretical predictions for $\alpha_1$ and $\chi_1$}
\label{theoretical}

As the scanning box glides along the surface, it frequently encloses high surface
steps created between neighboring grains. These are the box positions where the largest
height fluctuations are encountered, thus they give the main contribution to the roughness
(Eqs. \ref{defw} or \ref{defw1}).
If the box has size $r$ (i. e. $r$ pixels in
each direction), then the number of box positions that involve each high step is
proportional to $r$. 
Thus, the configurational average of Eq. (\ref{defw}) gives roughness $w$ proportional
to $r$. This gives $\alpha_1 =1$, as explained in
Ref. \protect\cite{grains} and confirmed by simulations of several models.

When Eq. (\ref{defw1}) is used, ${w'}^2$ is a configurational average.
The main contribution to that average also comes from box positions
enclosing high surface steps, thus, that average is proportional to $r$.
This gives $w'$ proportional to $r^{1/2}$, i. e., $\alpha_1=1/2$.

These results are confirmed by our simulations of
the RSOS model with cubic grains, as shown in Fig. 3a. It clearly shows the
remarkable difference in the scaling of $w$ and $w'$ for box sizes smaller than
the grain size, while the same exponent $\alpha_2$ after the crossover
represents the universality class of the process.

\begin{figure}[!t]
\includegraphics[width=7cm]{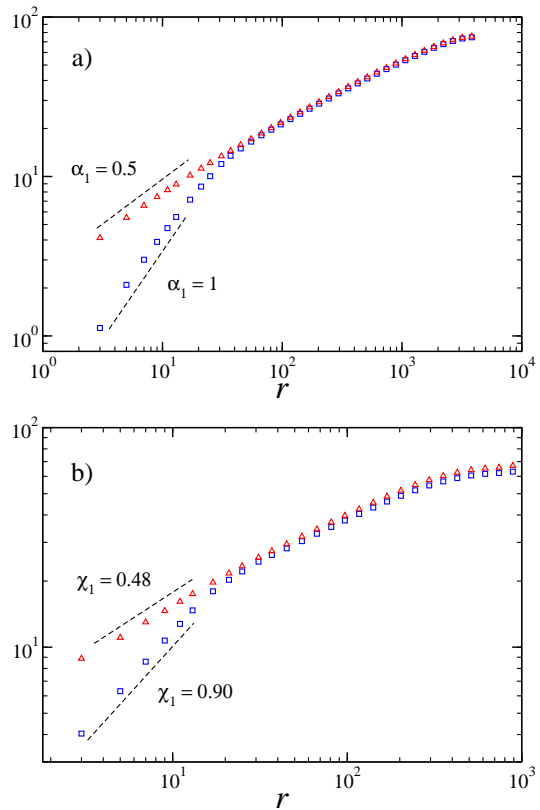}
\caption{Scaling with the window size $r$ of data for the KPZ model with cubic grains:
a) $w$ (blue squares) and $w'$ (red triangles); and b) $C'$ (blue squares) and $C$ (red triangles).}
\label{fig3}
\end{figure}

Similar situation is observed with the HHCF.
Again the main contribution for the configurational average comes from box positions
which involve high surface steps, thus this average is proportional to $r$. With
the most used definition of that function (Eq. \ref{defcorr}),
we have $C(r,t)$ proportional to $r^{1/2}$,
thus the crossover takes place with $\chi_1 =1/2$. Instead, if the scaling of
$C'(r,t)$ is analyzed, we expect $\chi_1=1$.

Simulations of the RSOS model with cubic grains show the predicted crossover,
as illustrated in Fig. 3b.
The exponent $\chi_1$ is very close to $1/2$ for $C(r,t)$
and slightly below $1$ for $C'(r,t)$. Again, the universal exponent
$\chi_2$ is obtained after the crossover; as expected, $\alpha_2\approx \chi_2$.

With the usual definitions of surface roughness ($w$ - Eq. \ref{defw}) and HHCF
($C$ - Eq. \ref{defcorr}), the roughness exponents measured
before the crossover ($\alpha_1$, $\chi_1$) are different. This contrasts with
the expected universality after the crossover ($\alpha_2\approx \chi_2$).
Our analysis show that those discrepancies are effects of
the grainy morphology and the calculation method, in particular the order of
calculation of square root and configurational average in Eqs. (\ref{defw})
and (\ref{defcorr}).

\section{Effects of grain shape}
\label{shape}

Rounding of the surface grains may lead to $\alpha_1$ between $0.85$ and $1$,
as shown in Ref. \protect\cite{grains}.
However, the replacement of the cubic grains by the rounded or sharp structures in Fig. 2,
with $h\geq l$, leads to much more drastic changes in the initial exponents of $w(r,t)$ and
$C(r,t)$.

This result is illustrated in Figs. 4a and 4b for films grown with the KPZ model
and pyramidal grains of height $h=64$: $\alpha_1$ decreases to $0.71$ and $\chi_1$
increases to $0.61$ (while $\alpha_2\approx 0.39$). Fig. 4b also shows the
formation of a plateau in the HHCF before the second scaling regime, which is 
characteristic of all sharp grains with large heights.

\begin{figure}[!b]
\includegraphics[width=7cm]{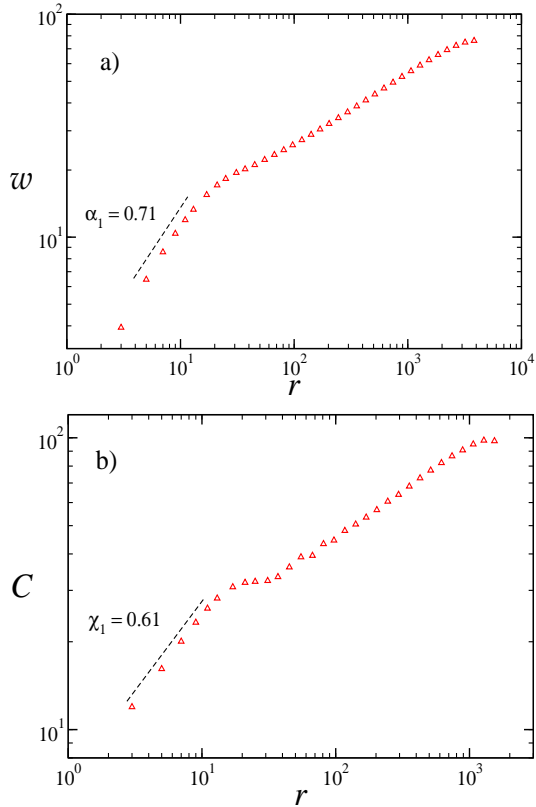}
\caption{a) Roughness ($w$) and b) HHCF ($C$) as a function of the window size $r$, for the KPZ
model with pyramidal grains of heigth $h=64$.}
\label{fig4}
\end{figure}

In Table I, we show the values of $\alpha_1$ and $\chi_1$ obtained for semi-elliptical,
conical and pyramidal grains with several heights.

A remarkable result is that films grown with the VLDS model have the same
exponents $\alpha_1$, $\chi_1$ up to the second decimal place, despite the
significant change in the asymptotic roughness exponent ($\alpha_2\approx 0.67$).
In Figs. 5a and 5b, we show results for conic grains
with height $h=32$, which give $\alpha_1= 0.809$ and $\chi_1=0.539$.
Those values are close to the KPZ values shown in Table I for the same grains.
The main differences from the models with KPZ scaling are that the
change in the slope of the roughness plot is smaller and there is a slope increase
in the HHCF plot when passing from the first to the second scaling regime.

\begin{figure}[!t]
\includegraphics[width=7cm]{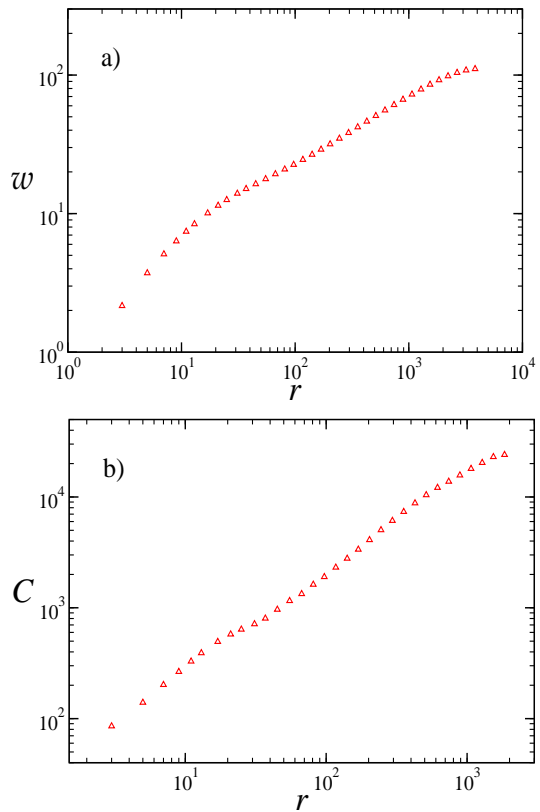}
\caption{a) Roughness ($w$) and b) HHCF ($C$) as a function of the window size $r$, for the VLDS
model with conical grains of heigth $h=32$.}
\label{fig5}
\end{figure}

Table I shows that $\alpha_1$ is much smaller than the limit $0.85$ obtained in
previous work \cite{grains} for many grain shapes, particularly for sharp conic and pyramidal
grains. With the structures studied here, the lower limit is
close to $0.71$, obtained with pyramidal grains. For $h\leq 3l$, the general trend is that
the increase of $h$ leads to decrease of $\alpha_1$. For larger $h$ (not shown in Table I),
a very slow increase of $\alpha_1$ towards $1$ is observed.

\begin{table}[!t]
\begin{center}
\begin{tabular}{cccccccc}
\hline\hline
$h$ &  &  $32$ &  &  $64$ & &  $96$ &  \\
\hline
$\alpha_{1}^{SE}$ &  & $0.826$ &  & $0.773$ &  & $0.763$   \\
 $\chi_{1}^{SE}$  &  & $0.504$ &  & $0.551$ &  & $0.589$  \\
\hline
$\alpha_{1}^C$ &  & $0.806$ &  & $0.768$ &  & $0.768$   \\
 $\chi_{1}^C$  &  & $0.539$ &  & $0.633$ &  & $0.694$  \\
\hline
$\alpha_{1}^P$ &  & $0.755$ &  & $0.710$ &  & $0.708$   \\
 $\chi_{1}^P$  &  & $0.535$ &  & $0.606$ &  & $0.645$  \\
\hline\hline
\end{tabular}
\caption{Exponents obtained from $w$ ($\alpha_1$) and $C$ ($\chi_1$) in KPZ films with
semi-elliptical (SE), conical (C) and piramidal (P) grains.}
\label{table1}
\end{center}
\end{table}

The relative changes in $\chi_1$ are much larger, attaining almost $40 \%$
for conic grains with $h=3l$ (see Table I). Indeed, this is the grain shape that provides higher
deviations from the flat grain value $\chi_1 =0.5$. A monotonic increase of $\chi_1$
is observed when taller grains are studied.

The above results show that sharp grain shapes bring closer the exponents
$\alpha_1$ and $\chi_1$, in contrast with the very different values for flat grains
($1$ and $0.5$, respectively - Sec. \ref{theoretical}).
In some cases, they are surprisingly close; for instance,
they differ only $10\%$ for conic grains with $h=3l$.

In Table II, we show exponents $\alpha_1$ and $\chi_1$ obtained from
the scaling of $w'(r,t)$ and $C'(r,t)$. They should be compared
with the respective flat grain values $0.5$ and $1$.

Comparison of results in Tables I and II show that sharp grain shapes also bring
closer the values of $\alpha_1$ measured from $w$ and $w'$ scaling, which are
very different for flat grains ($1$ and $0.5$, respectively - Sec. \ref{theoretical}).
It is particularly interesting to observe that $\alpha_1$ differs only $3\%$
when calculated from $w$ or $w'$ in films with pyramidal shapes with $h=3l$.
These values may be incorrectly interpreted as true roughness exponents because the
same $\alpha_2$ is expected for $w$ and $w'$. This type of erroneous interpretation
can be  avoided if one accounts for the effects of a wide range of
grain shapes and sizes and investigates other quantities, such as HHCF.

\begin{figure}[!b]
\includegraphics[width=8cm]{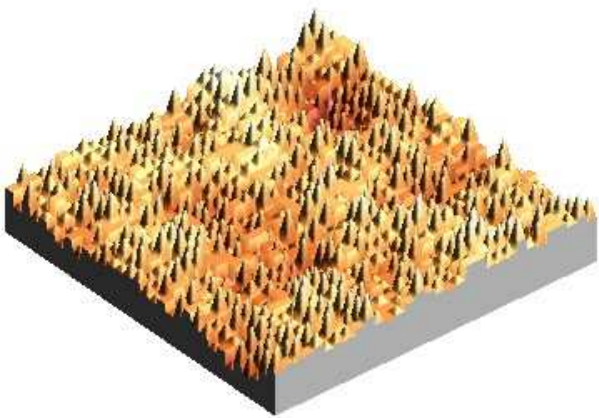}
\caption{Film surface with $1/4$ of the grains flat and $3/4$ pyramidal with heights $h=32$,
$h=64$, and $h=96$ equally distributed.}
\label{fig6}
\end{figure}

The crossover size $r_c$ is defined at the intersection of the linear fits of the initial regime
and the second scaling regime of roughness or HHCF, as illustrated in Fig. 1..
Despite the wide range of values of $\alpha_1$ and $\chi_1$ shown in Tables I and II,
a remarkable result is that $r_c$ is always very close
to the grain size $l$, for KPZ and VLDS models.
Using $l=32$, our estimates range between $r_c=30$ and $r_c=34$,
which corresponds to a maximum difference of $7\%$. Consequently, $r_c$ can always be used as a
reliable estimate of the grain size.

In the above models, we considered surfaces with uniform grain height. However, we
also analyzed the effect of distributions of grain heights, since this is the situation
in real surfaces. In all cases,
we observe that the exponents $\alpha_1$ and $\chi_1$ are near the averages of those
obtained with a single value of grain height.

\begin{table}[!t]
\begin{center}
\begin{tabular}{cccccccc}
\hline\hline
$h$ &  &  $32$ &  &  $64$ & &  $96$ &  \\
\hline
$\alpha_{1}^{SE}$ &  & $0.523$ &  & $0.576$ &  & $0.621$   \\
 $\chi_{1}^{SE}$  &  & $0.754$ &  & $0.711$ &  & $0.705$  \\
\hline
$\alpha_{1}^C$ &  & $0.554$ &  & $0.650$ &  & $0.719$   \\
 $\chi_{1}^C$  &  & $0.712$ &  & $0.693$ &  & $0.710$  \\
\hline
$\alpha_{1}^P$ &  & $0.556$ &  & $0.638$ &  & $0.687$   \\
 $\chi_{1}^P$  &  & $0.664$ &  & $0.633$ &  & $0.640$  \\
\hline\hline
\end{tabular}
\caption{Exponents obtained from $w'$ ($\alpha_1$) and $C'$ ($\chi_1$) in KPZ films with
semi-elliptical (SE), conical (C) and piramidal (P) grains.}
\label{table2}
\end{center}
\end{table}

An example of a film surface with such random grain distribution is shown in Fig. 6:
$1/4$ of the grains are flat and $3/4$ have pyramidal shape, with equally distributed heights $h=32$,
$h=64$, and $h=96$. For that surface, we obtain $\alpha_1=0.742$ and $\chi_1=0.554$,
which is close to the average of the results in Table I for those shapes.

\section{Comparison with experimental results}
\label{experimental}

In the experimental works discussed below, the exponent $\alpha_1$ defined here is
frequently named local roughness exponent $\alpha_{loc}$, as a reference to the small
lengthscale behavior and/or to a possible anomalous scaling.

Several experimental works have already shown the crossover of Fig. 1 with $\alpha \approx 1$,
which is explained by the growth models with flat or slightly rounded grains \cite{grains}.
Among those works, we highlight the study of rf sputtered
$LiCoO_x$ films by Kleinke et al \cite{kleinke1999}, which gives $0.91\leq \alpha_1\leq 0.95$;
the spray pyrolysis growth of $ZnO$ films by Eboth\'e et al \cite{ebothe2003}, which gives
$0.94\leq \alpha_1\leq 0.97$ for high flow rates; the electrodeposition of cooper by
Mendez et al \cite{mendez} and of gold by V\'azquez et al \cite{vazquez},
which give $\alpha_1=0.87\pm 0.06$
and $\alpha_1=0.90\pm 0.06$, respectively; the electrochemical
roughening of silver electrodes by Otsuka and Iwasaki \cite{otsuka}, which gives 
$\alpha_1$ between $0.95$ and $0.98$; and the pulsed laser deposition of 
$La$ modiﬁed-$PbTiO_3$ films of Vasco et al \cite{vasco}, which have $\alpha_1 = 1$.

However, many works show the same crossover with exponents $\alpha_1$
between $0.7$ and $0.85$, and surface images confirm the presence of grains
of approximately conic or pyramidal shape, much higher than the
steps between neighboring grains. These features are observed in films
of various materials and substrates, deposited with different techniques. This justifies
our approach with geometrical models, independently of the particular growth
dynamics.

Among the applications to inorganic materials, we find some vapor deposited gold films
by Vazquez et al, which have $\alpha_1\approx 0.83$ - see Fig. 1c and Fig. 3 of Ref.
\protect\cite{vazquezSS}. One of the niquel oxide film samples deposited by sputtering
in Ref. \protect\cite{tersio} have $\alpha_1 =0.70$, and the AFM image show the
qualitative features of our models with sharp grainy structure. Nearly the same
exponent ($\alpha_1=0.71$) is obtained with $Ni$ films electrodeposited on
indium tin oxide substrates in Refs. \protect\cite{mendomeJCG,mendomeMCP}.
Several $Ni-Zn$ alloy films of 
Ref. \protect\cite{hiane} show the crossover in roughness scaling, with most estimates
of $\alpha_1$ in the range $[0.80,0.83]$. This is consistent with
our models of semi-elliptical grains of lower $h$, and the images actually show a
smooth grain morphology.

The same features are also observed in organic materials. Films formed
with bilayers of poly(allylamine hydrochloride) and a side-chain-substituted azobenzene
copolymer (Ma-co-DR13), after deposition of $10$ or $20$ bilayers, show grains with a
broad size distribution, and the
initial roughness exponents $0.81$ and $0.79$ \cite{nara2005}.
AFM images of chemically deposited polyaniline thin films on glass substrates \cite{nara2007}
have similar features, but, as far as we know, roughness scaling was not studied
with those images. The surface of Langmuir–Blodgett films of polyaniline and a
neutral biphosphinic ruthenium complex (Rupy) of Ref. \protect\cite{naraNano2007}
also show those grainy features with
some high peaks, and initial roughness exponents are in the range
$0.66\leq \alpha_1\leq 0.81$ for thicknesses between $1$ and $21$ layers. However,
most estimates of $\alpha_1$ are between $0.72$ and $0.76$ \cite{naraNano2007},
in good agreement with our results for pyramidal grains.

It is interesting to observe that
some surface images shown in Refs. \protect\cite{mendomeJCG,mendomeMCP,nara2005,nara2007,naraNano2007}
have features similar to the model illustration in Fig. 6 (in most cases without the flat grains).
This comparison reinforces our interpretation of the exponents measured in those works.

Similar results are obtained in etching of silicon surfaces
in Ref. \protect\cite{marta2002}: $\alpha_1$ is found between $0.70$ and $0.87$ when the
$(111)$ surface is etched by an $NaOH$ solution in contact with a non-saturated aqueous
environment.

It is also important to recall that there are systems with crossover in the roughness
scaling which do not show the sharp grainy features of our models, and consequently
deserve separate investigation. For instance, the images of another sample from
Ref. \protect\cite{tersio} does not show those features, but the roughness shows a
slow crossover with $\alpha_1=0.52$. Pyroclastic deposits of Mt. Etna show
roughness scaling crossover with $\alpha_1$ between $0.47$ and $0.67$,
but the images do not support modeling by grainy structures \cite{mazzarini}..
There are also systems with sharp grainy structures and small $\alpha_1$, such
as some $Ni$ films of Ref. \protect\cite{mendomeMCP} ($0.55\leq\alpha_1\leq 0.61$),
which also would deserve a separate investigation (those films have
$0.12\leq\alpha_2\leq 0.22$, which also cannot be easily explained with the well known
kinetic growth theories \cite{barabasi}).

The crossover in HHCF scaling obtained in some systems can also be related to our models.
For instance, Manes et al \cite{manes} used HHCF as a measure of fresh snow roughness and
obtained $\chi_1$ between $0.58$ and $0.62$ in a set of five experiments. These values
are consistent with our model with very high semi-ellipsoidal grains ($h=3l$) or with
conic or pyramidal grains with $h=2l$ or less.

Again, there are also systems where a crossover of HHCF scaling is observed but whose
images do not show the features of our models. An example is Ref. \protect\cite{martasergio},
where $\chi_1 =0.84$ was obtained for paraphin films deposited on stainless steel covered
with amorphous carbon.

Results of the recent work on pentacene island growth on stepped oxide surfaces \cite{conrad} can also
be related to our models. First, for long lengths, the HHCF has exponent $2\chi =1$, which
is expected for height fluctuations dominated by the surface steps; indeed, arguments analogous to those
for flat grains (Sec. \ref{theoretical}) give $\chi =1/2$. However, for small lengths, height
fluctuations in the surface terraces (due to pentacene islands) lead to the increase of
the HHCF exponent to the range $[0.69,0.8]$. Recent works showing evidence of anomalous scaling
in organic and inorganic film surfaces
also give estimates of HHCF exponents above $1/2$ at short lengthscales \cite{kim,zhang}. Although
both short and long range dynamics may be much more complex than in our models, a simple geometric
interpretation of the short range exponents may also be considered due to the presence of grainy
structures in the  surface images.

\section{Conclusion}
\label{conclusion}

We extended the work on growth models with grainy surfaces to analyze the
effects of the grain shape, of the method of calculation of
averages of squared quantities, of the working quantity (roughness or HHCF) and
of the universality class of the growth process.
For all models, grain shapes, and methods of analysis, we observe
crossovers at box sizes very close to the average grain size.
We also show that a very broad range of the initial exponent $\alpha_1$ is found
for the roughness scaling, decreasing from $1$ for flat grains to $0.71$ for some sharp pyramidal
grains. The initial exponent $\chi_1$ of HHCF scaling increases from approximately
$0.5$ for flat grains to values larger than $0.7$ for sharp conic grains.
Simulations of KPZ and VLDS models show that the universality class has no significant
effect on the estimates of $\alpha_1$ and $\chi_1$.
The range of $\alpha_1$ presented here explains results of some recent experimental
works with different materials and deposition methods. This gives additional support
to the geometric interpretation of the crossover in roughness scaling for a
variety of systems.

\acknowledgments

FDAAR acknowledges support from CNPq and Faperj (Brazilian agencies).




\end{document}